# TARGETED SELF SUPERVISION FOR CLASSIFICATION ON A SMALL COVID-19 CT SCAN DATASET

*Nicolas Ewen[*], Naimul Khan[†]*

[*] Data Science, Ryerson University, Toronto, ON
[†] Electrical, Computer, and Biomedical Engineering, Ryerson University, Toronto, ON

## ABSTRACT

Traditionally, convolutional neural networks need large amounts of data labelled by humans to train. Self supervision has been proposed as a method of dealing with small amounts of labelled data. The aim of this study is to determine whether self supervision can increase classification performance on a small COVID-19 CT scan dataset. This study also aims to determine whether the proposed self supervision strategy, targeted self supervision, is a viable option for a COVID-19 imaging dataset. A total of 10 experiments are run comparing the classification performance of the proposed method of self supervision with different amounts of data. The experiments run with the proposed self supervision strategy perform significantly better than their non-self supervised counterparts. We get almost 8% increase in accuracy with full self supervision when compared to no self supervision. The results suggest that self supervision can improve classification performance on a small COVID-19 CT scan dataset. Code for targeted self supervision can be found at this link: https://github.com/Mewtwo/Targeted-Self-Supervision/tree/main/COVID-CT

## 1. INTRODUCTION

The objective of this study is to determine whether self supervision can help improve classification performance of convolutional neural networks on a small COVID-19 CT scan dataset [1]. A new strategy for self supervision is proposed, and its performance is compared to a baseline.

Some small medical imaging datasets can be hard to train convolutional neural network (CNN) models on. These models require large amounts of annotated data, which is often not publicly available [2][3]. Methods addressing this issue include transfer learning, and data augmentation [2][3][4].

A number of previous works that have dealt with classification on small medical imaging datasets have found success using convolutional neural networks [5][2][3][6][1][7]. Some papers have tried to train networks from scratch using these small datasets [5][7], but transfer learning from a network pre-trained on ImageNet often produces better results [6][7]. A number of recent works showed that performing self supervised learning on the networks pre-trained using ImageNet prior to performing transfer learning can boost the performance of the classifiers [8][9][10].

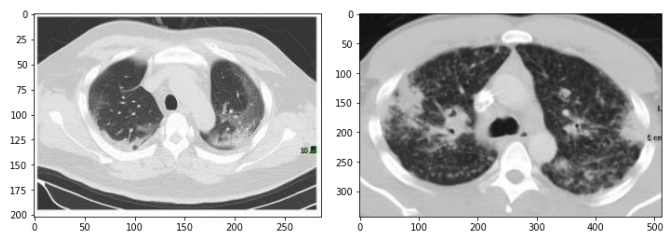

**Fig. 1**. Example of a CT scan with Covid-19 (left), and without (right).

Transfer learning is a technique where a network is pre-trained for a task on a separate dataset, and is then fine tuned for the target task, using the target dataset [7]. It allows the target task to benefit from the training on the unrelated task. Often in CNNs, the early layers of the network are responsible for simple and reusable features, and are therefore often useful for different tasks [11]. It is useful when dealing with limited training data, since the network will not have to thoroughly train early layers. This reduces how much training is needed, and how much data is needed to train the network [7]. Self supervised learning is a form of unsupervised learn- ing where the network is pre-trained using automatically labelled data, or unlabeled data. After the pre-training is performed, the network is fine tuned using transfer learning onto the target dataset. A pre-text task is chosen along with a dataset related to the target dataset for the pre-training so that in learning to solve this task, the network can learn important feature semantics [8]. An advantage of self supervised learning is that it allows the network to learn important visual features for the target task [8]. While transfer learning helps learn simple and reusable features in the early layers, it can be limited to only helping in early layers. Transfer learning can be performed from different image types, with different content. For example, from ImageNet data to medical imaging, there is a large domain shift [7]. Performing self supervised learning using data closely related to the target data can allow more of the network to be pre-trained. It can allow the network to get used to looking for specific image semantics [8]. Self supervision is useful when there is a limited amount of labelled training data. Since the network will be pre-trained to identify the target image semantics, it will not need as much labelled training data to be trained thoroughly. It can be very effective when there is a large amount of unlabelled data available [9].

Previous works that performed self supervision describe what they think a good self supervision strategy should do [9]. For example, Chen et al. argue that the self supervision should learn features that capture image semantics, be useful for different kinds of tasks, and should be simple [9]. They also argue that it is good to design the CNN architecture in such a way that the target tasks can take full advantage of the pre-trained weights [9].

Our main contribution in this work is to highlight and demonstrate a simple self supervision strategy. While many self supervision strategies are designed for a variety of downstream tasks, our proposed *targeted self supervision* is designed around the downstream task, and is only concerned with performance on the specific downstream task. Many self supervision strategies will remove some of the top layers after the self supervision step, and replace them with layers for a chosen downstream task. In targeted self supervision, we start with the downstream task, and perform the self supervision with the layers chosen for the downstream task. In this way, all the layers of the network can benefit from the self supervision. The pre-text task is designed with the specific downstream task in mind. This can allow better pre-training during the self supervision step. We demonstrate one example of this strategy, designing a pre-text task that fits well with the chosen architecture for the downstream task. We offer insight on how targeted self supervision can be used and implemented to boost performance. We perform 10 experiments demonstrating that targeted self supervision is advantageous. We produce competitive results in our experiments, showing that gradually increasing targeted self-supervision results in significant increase in performance.

## 2. METHODOLOGY

### 2.1. Dataset

The dataset that we used for this paper was the COVID-CT dataset, a dataset of chest CT scan images, created by Zhao et al. [1]. This dataset contains 349 COVID-19 positive CT scans from 216 patients, as well as 397 COVID-19 negative CT scans [1]. The images were collected from publicly available preprints from medRxiv and bioRxiv [1]. This gives a total of 746 CT scans. Early on during the COVID-19 pandemic, this dataset was one of the largest publicly available COVID-19 CT scan datasets [1]. See Fig. 1 for some examples of the CT scans from the dataset. Only images that had a caption indicating whether it was a case with Covid-19, or a healthy patient were taken. The training set has 543 images, and the test set has 203 images. Even though this dataset was one of the largest, it is still fairly small. This was because there was a limited number of publicly available COVID-19 CT scans, and because it can be expensive to have medical professionals annotate the images by hand [7].

### 2.2. Pre-processing

The Covid-CT dataset has pre-arranged training, validation, and test sets. The pre-arranged sets are used for this project. Since some images came from the same patients [1], we did not want to make our own randomized splits, to avoid data leakage. The validation set was merged with the training set to create a larger training set. Therefore, hyperparameter tuning was not part of this experiment. No further pre-processing was done for this dataset.

### 2.3. Self Supervision

In this paper, a new method of self supervision is proposed, called targeted self supervision. Some potential advantages of this method are that it can be simple, can have low computing cost, and can be used to pre-train the entire network. Other self supervision methods can have high computing cost [7][9][10], can be more difficult to implement [9], or only train part of the network [9]. A potential drawback of this method is that it requires the user to think of a suitable pre-training task.

The proposed method is as follows: Once the architecture and target task have been decided, a related pretext task that can use the exact same architecture is found, and the network is pre-trained for this related pretext task. This allows the entire network to be pre-trained. A simple related pretext task can be chosen to pre-train the network to keep cost low.

Binary classification is being performed on this dataset. Therefore, the related pretext task should also be binary classification to take advantage of the entire network pre-training.

For the Covid-CT dataset, a new dataset was created by flipping all images in the dataset horizontally and saving them. This doubled the number of images that could be used to perform self supervision. The binary classification is determining whether an image is a flipped image or not. This is a good self supervision task because the images used will have the same image qualities as those in the target dataset [9], and since the images are fairly similar (see examples in Fig. 2), the network will need to learn to inspect and understand the body. For example, it may try to determine which lung is larger to tell if the image is flipped. This pretext task is simple to run, and should not take a huge amount of resources.

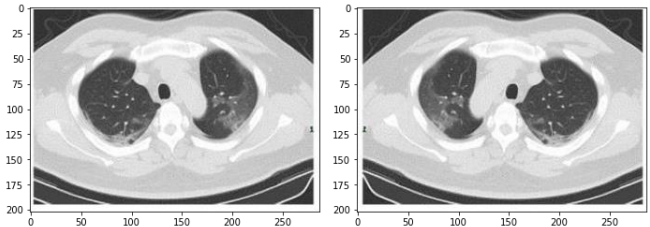

**Fig. 2**. Example of a CT scan (left) and the flipped counterpart (right).

### 2.4. Dataset Creation

To create the dataset for the self supervision, all the images were first saved to a new folder as JPEGs. The training and test sets were still kept separate. For each image, a new image was created by flipping it horizontally, as seen in Fig. 2. These new flipped images were then saved to a separate folder as JPEGs. The result of creating the dataset in this way is that there are an equal number of images in the original folder and in the flipped folder.

Some of the experiments for this paper only used part of the data for the self supervision step. For these experiments, partial datasets were created. The images in the original folder and the flipped folder were in the same order, so a fraction of each of the folders was taken. Three partial dataset were created. They contained 25 percent, 50 percent, and 75 percent of the data respectively. The data was not selected randomly, but sequentially. This means that all of the data in the 25 percent dataset is also contained in the 50 percent dataset, and so on.

### 2.5. Experimental Set Up

In order to test the proposed method of self supervision, 10 experiments were run. Half of these experiments were run with a DenseNet CNN base [12], and the other half were run with an InceptionV3 CNN base [13]. For each of the CNN bases, five experiments were run.

In each experiment, a different amount of training data was used for self supervision. In the first experiment for each CNN base, no self supervision is performed. These two runs are only transfer learning, and were used as a baseline.

In the second experiment for each CNN base, 25 percent of the data is used for the self supervision step. Transfer learning is then performed using the original dataset, and all the data is used. The third, fourth, and fifth experiments for each CNN base are similar, except that 50 percent, 75 percent, and 100 percent of the data is used for the self supervision step, respectively. See Table 1 for a breakdown of the experiments.

All CNN bases were initialized from models trained on ImageNet, and all dense tops were randomly initialized.

In the self supervision step there are three training runs total. The first run is to train the dense top that was randomly initialized. We used the RMSProp optimizer, with a learning rate of 0.0001, for 30 epochs. A callback was added to save the model with the lowest loss. The model that was saved is then loaded before starting the second run. The second run trains the entire network, not just the dense top. We used stochastic gradient descent with mini batches of size 4, and a learning rate of 0.00001 for 30 epochs. A callback was again used to save the model with the lowest loss, which was then loaded before starting the third run. The third run was set up in the same way as the second, except that the learning rate was reduced to 0.000001.

**Table 1**. Experiments

|  | Proposed Experiments | |
| --- | --- | --- |
|  | *CNN Base* | *Fraction for Self Supervision* |
| Exp. 1 | InceptionV3 | 0.0 |
| Exp. 2 | InceptionV3 | 0.25 |
| Exp. 3 | InceptionV3 | 0.5 |
| Exp. 4 | InceptionV3 | 0.75 |
| Exp. 5 | InceptionV3 | 1.0 |
| Exp. 6 | DenseNet169 | 0.0 |
| Exp. 7 | DenseNet169 | 0.25 |
| Exp. 8 | DenseNet169 | 0.5 |
| Exp. 9 | DenseNet169 | 0.75 |
| Exp. 10 | DenseNet169 | 1.0 |

The transfer learning step is the same for all experiments. After the self supervision, we then transferred onto the target task using the original dataset without the flipped images. All of the original dataset was used. For this step there were two training runs total. The first run only trained the dense top. We used the RMSProp optimizer, with a learning rate of 0.0001, for 30 epochs. The second run then trained the entire network. We used stochastic gradient descent with mini batches of size 4, and a learning rate of 0.00001 for 30 epochs. A callback was used to save the model with the highest accuracy from either of the two runs, and the model that was saved was used as the final model. See Fig. 1 for examples of images used in this step.

### 2.6. Experiments

A number of previous works have produced good results on small medical imaging datasets with two dense layers on top [5][6][7]. To reduce the computing cost, a single dense layer of size 1024 was used instead, with a sigmoid activation on top. Data augmentation was also not used to reduce computing cost. As was done in [1], mini batches of size four were used for the SGD. Previous papers report good results when using DenseNet169 and InceptionV3 as the convolutional bases [7][3][2]. Experiments were run with both architectures. For both architecture bases, the final training for the target task was always the same. The only difference was how much data was used in the self supervision step. This gave a total of ten experiments to run. The resulting model from each experiment was measured using three performance measures on the test set: the accuracy of the model, the area under the receiver operating curve (AUC), and the F1 score.

## 3. RESULTS

Table 2 shows the results of the experiments. The performance measures used were accuracy, AUC, and the F1 score.

The results are promising. All experiments that included a self supervision step show improved performance in all three metrics over experiments with only transfer learning.

The highest performing model came from experiment ten, with a DenseNet169 CNN base, and the full dataset was used for the self supervision. The model achieved a test set accuracy of 0.8621, an AUC of 0.8609, and an F1-score of 0.8704. To the best of our knowledge, this model achieves the best F1-score in the literature as of the time of this writing. This suggests that our proposed method of self supervision can produce competitive results, even though it is simple and cheap.

Interestingly, the top performing model with an inceptionV3 base came from experiment three, which used only 50 percent of the data for the self supervision step. This model was the second highest performing model overall. We can think of two possible explanations for this. One possibility is that the 50 percent dataset contained images that were particularly good for the model to learn appropriate representations. This could explain why for both the InceptionV3 and the DenseNet169 bases, the models from the experiments that used 50 percent of the data outperformed the models from the experiments that used 75 percent of the data for the self supervision step. Another possible explanation is that the InceptionV3 base could learn better representations with less data than the DenseNet169 base, while the DenseNet169 base could learn better representations with more data than the InceptionV3 base. This could explain why the top performing model with a DenseNet169 base comes from experiment ten, which used 100 percent of the data for self supervision, while the top performing model with an InceptionV3 base comes from experiment three, which used 50 percent of the data.

For the Covid-CT dataset, to the best of our knowledge, the top performance metrics reported are: accuracy of 0.8834 [14], AUC of 0.94 [7], and an F1 score of 0.8670 [14] . Without hyperparameter tuning or data augmentation, one of the self supervision runs achieved performance metrics of: accuracy of 0.8621, AUC of 0.8609, and an F1 score of 0.8704. The accuracy and F1 score achieved are competitive with top results so far, with the F1 score being the current top score. This also suggests that a study that includes hyperparameter tuning and data augmentation can possibly produce top performance results on other performance metrics on this dataset.

## 4. CONCLUSIONS

In this paper, we propose a new method for self supervision that we call *targeted self supervision*. Unlike regular self supervision where a pretext task is designed for different downstream tasks, we start with the downstream task in targeted self supervision. Unlike regular self supervision where the top layers are removed for different downstream tasks, we use the same network for pre-text and downstream tasks, thus simplifying the process of self-supervision.

Given the results of the experiments, we conclude that self supervision can help increase model performance in the diagnosis of COVID-19 CT scans over transfer learning from models pre-trained on ImageNet. Targeted self supervision can produce competitive models, while being simple to perform, and cheap to run.

While our models produce good results, especially for the F1-score, we did not perform any hyperparameter tuning due to lack of computational resources. A potential future area of research could be to tune hyperparameters, and find more optimal settings before performing targeted self supervision. This could potentially produce models with greater performance. Another area of future research that could boost model performance is data augmentation, which was not done for our experiments. From our literature review, we believe that with data augmentation, the models could have achieved higher performances.

Another potential area of future research could be to test different divisions of the data for the self supervision. Such a study could further clarify why the highest performing model with an InceptionV3 base was the one where the self supervision step only used 50 percent of the data.

Future research can also focus on expanding the testing of targeted self supervision to other datasets, and testing other pretext tasks for targeted self supervision.

**Table 2**. Results of Experiments

| | **Results** | | |
| --- | --- | --- | --- |
| | *Accuracy* [a] | *AUC* | *F1 Score* |
| Exp. 1 | 0.7882 +/- 0.0562 | 0.7850 | 0.8106 |
| Exp. 2 | 0.8276 +/- 0.0520 | 0.8241 | 0.8472 |
| Exp. 3 | 0.8621 +/- 0.0474 | 0.8612 | 0.8692 |
| Exp. 4 | 0.8423 +/- 0.0501 | 0.8415 | 0.8505 |
| Exp. 5 | 0.8473 +/- 0.0494 | 0.8469 | 0.8531 |
| Exp. 6 | 0.7882 +/- 0.0562 | 0.7888 | 0.7902 |
| Exp. 7 | 0.8227 +/- 0.0525 | 0.8214 | 0.8333 |
| Exp. 8 | 0.8424 +/- 0.0501 | 0.8408 | 0.8532 |
| Exp. 9 | 0.8276 +/- 0.0520 | 0.8279 | 0.8309 |
| Exp. 10 | 0.8621 +/- 0.0474 | 0.8609 | 0.8704 |

[a]Accuracy with a 95 percent confidence interval.

## 5. REFERENCES

[1] Xingyi Yang, Xuehai He, Jinyu Zhao, Yichen Zhang, Shanghang Zhang, and Pengtao Xie, "Covid-ct-dataset: A ct scan dataset about covid-19," 2020.

[2] Richa Agarwal, Oliver Diaz, Xavier Lladó, Moi Hoon Yap, and Robert Martí, "Automatic mass detection in mammograms using deep convolutional neural networks," *Journal of Medical Imaging*, vol. 6, no. 3, pp. 1 – 9, 2019.

[3] Hiba Chougrad, Hamid Zouaki, and Omar Alheyane, "Convolutional neural networks for breast cancer screening: Transfer learning with exponential decay," *CoRR*, vol. abs/1711.10752, 2017.

[4] N. Khalifa M. Loey, G. Manogaran, "A deep transfer learning model with classical data augmentation and cgan to detect covid-19 from chest ct radiography digital images," 2020.

[5] Andrea Duggento, Marco Aiello, Carlo Cavaliere, Giuseppe Leonardo Cascella, Davide Cascella, Giovanni Conte, Maria Guerrisi, and Nicola Toschi, "An ad hoc random initialization deep neural network architecture for discriminating malignant breast cancer lesions in mammographic images," *Contrast Media & Molecular Imaging*, vol. 2019, pp. 1–9, 05 2019.

[6] Dina Ragab, Maha Sharkas, Stephen Marshall, and Jinchang Ren, "Breast cancer detection using deep convolutional neural networks and support vector machines," *PeerJ*, vol. 7, pp. e6201, 01 2019.

[7] Xuehai He, Xingyi Yang, Shanghang Zhang, Jinyu Zhao, Yichen Zhang, Eric Xing, and Pengtao Xie, "Sample-efficient deep learning for covid-19 diagnosis based on ct scans," *medRxiv*, 2020.

[8] Longlong Jing and Yingli Tian, "Self-supervised visual feature learning with deep neural networks: A survey," *CoRR*, vol. abs/1902.06162, 2019.

[9] L Chen, P Bentley, K Mori, K Misawa, M Fujiwara, and D Rueckert, "Self-supervised learning for medical image analysis using image context restoration," *Medical Image Analysis*, vol. 58, pp. 1–12, 2019.

[10] Kaiming He, Haoqi Fan, Yuxin Wu, Saining Xie, and Ross Girshick, "Momentum contrast for unsupervised visual representation learning," 2020.

[11] Francois Chollet, *Deep Learning with Python*, Manning Publications Co., USA, 1st edition, 2017.

[12] Gao Huang, Zhuang Liu, and Kilian Q. Weinberger, "Densely connected convolutional networks," *CoRR*, vol. abs/1608.06993, 2016.

[13] C. Szegedy, Wei Liu, Yangqing Jia, P. Sermanet, S. Reed, D. Anguelov, D. Erhan, V. Vanhoucke, and A. Rabinovich, "Going deeper with convolutions," in *2015 IEEE Conference on Computer Vision and Pattern Recognition (CVPR)*, 2015, pp. 1–9.

[14] Arnab Mishra, Sujit Das, Pinki Roy, and Sivaji Bandyopadhyay, "Identifying covid19 from chest ct images: A deep convolutional neural networks based approach," *Journal of Healthcare Engineering*, vol. 2020, pp. 1–7, 08 2020.